\documentclass[12pt, a4paper, reqno]{amsart}
\usepackage{mathrsfs}
\usepackage{bbm}
\usepackage{amsmath,amscd,amssymb,latexsym}
\usepackage[all]{xy}

\input xypic

\evensidemargin=\oddsidemargin
\DeclareMathVersion{can}
\DeclareMathAlphabet{\can}{OT1}{cmss}{m}{n}
\newtheorem{thm}{Theorem}[section]
\newtheorem{cor}[thm]{Corollary}
\newtheorem{lem}[thm]{Lemma}

\newtheorem{rem}[thm]{Remark}
\newtheorem{exa}[thm]{Example}
\theoremstyle{definition}
\newtheorem{defn}[thm]{Definition}
\theoremstyle{fact}

\theoremstyle{conjecture}

\numberwithin{equation}{section}

\newcommand{\ord}{\operatorname{ord}}

\begin{document}
\title[ Self-reciprocal and self-conjugate-reciprocal polynomials] { Self-reciprocal and self-conjugate-reciprocal irreducible factors of $x^n-\lambda$ and their applications}

\author[Y. Wu]{Yansheng Wu}
\address{\rm Department of Mathematics, Nanjing University of Aeronautics and Astronautics,
Nanjing, 211100, P. R. China; State Key Laboratory of Cryptology, P. O. Box 5159, Beijing, 100878, P. R. China; Department of Mathematics, Ewha Womans University, Seoul, 03760, South Korea}
 \email{wysasd@163.com}

\author[Q. Yue]{Qin Yue}
\address{\rm Department of Mathematics, Nanjing University of Aeronautics and Astronautics,
Nanjing, 211100, P. R. China; State Key Laboratory of Cryptology, P. O. Box 5159, Beijing, 100878, P. R. China}
\email{yueqin@nuaa.edu.cn}

\author[S. Fan]{Shuqin Fan}
 \address{\rm State Key Laboratory of Cryptology, P. O. Box 5159, Beijing, 100878, P. R. China}
\email{fansq@sklc.org}




\begin{abstract} In this paper, we present some necessary and sufficient conditions under which an irreducible polynomial is self-reciprocal (SR) or self-conjugate-reciprocal (SCR). By  these characterizations, we obtain some enumeration formulas   of SR and SCR  irreducible factors of $x^n-\lambda$, $\lambda\in \Bbb F_q^*$, over $\Bbb F_q$,   which are just  open questions posed  by  Boripan {\em et al} (2019). We also count the numbers of Euclidean and Hermitian LCD constacyclic codes and show some well-known results on Euclidean and Hermitian self-dual constacyclic codes in a simple and direct way.
\end{abstract}

\maketitle

\section{Introduction}

 In coding theory, cyclic codes have been extensively investigated  due to  efficient error detection and correction. 
Self-reciprocal (SR) polynomials  were first used to characterize when  a cyclic code is reversible by  Massey \cite{M} in 1964. Since then,  some properties of SR polynomials are widely studied, see \cite{ F, FR, G1,  GK, G2, GB,  HB, YM2}. They also have applications in many areas of research, in particular in coding
theory, see Euclidean linear complementary dual (LCD)  cyclic codes \cite{L1, LDL, YM1}  and Euclidean self-dual cyclic codes \cite{CLZ, JLX}.


 Constacyclic codes is a well-known generalization of cyclic  codes. SR polynomials can be also used to investigate Euclidean  LCD and self-dual constacyclic codes, see \cite{WY}.
Next let us introduce the definition of constacyclic codes over  finite fields. Let $\Bbb F_q$ be a finite field of order $q$, where $q$ is a power of some prime $p$.
An $[n, k, d]$ linear code $\mathcal{C}$ over $\Bbb F_q$ is a $k$-dimensional subspace of $\Bbb F_q^n$ with minimum  Hamming  distance $d$.
For $\lambda\in \Bbb F_q^*$, the ring $\frac{\Bbb F_q[x]}{\langle x^n-\lambda \rangle}$ is a principle ideal ring, i.e., every ideal can be generated by a monic divisor of $x^n-\lambda$.
  A $\lambda$-constacyclic code of length $n$ over $\Bbb F_q$ can be viewed as an ideal  $\left\langle g(x)\right\rangle$ in the ring $\frac{\Bbb F_q[x]}{\langle x^n-\lambda\rangle}$, where  $g(x)\mid  (x^n-\lambda)$. Hence the irreducible factorization of $x^n-\lambda$ in $\Bbb F_q[x]$ determines all $\lambda$-constacyclic codes of length $n$. When $p\nmid   n$, a $\lambda$-constacyclic code is called a simple-root constacyclic code. When $p\mid n$, a $\lambda$-constacyclic code is called a repeated-root constacyclic code. When $\lambda=1$ and $\lambda=-1$,  $\lambda$-constacyclic codes are known as cyclic codes and negacyclic codes, respectively. For more details about constacyclic codes, the reader is referred to \cite{CDL,CDL2,CDLW, CFLL,D1, D3, D5,D7,LM,LLKZ, WY, ZTG}.

 Most recently, Boripan {\em et al.} \cite {BJU} introduced the concept of self-conjugate-reciprocal (SCR) polynomials over $\Bbb F_{q^2}$. They gave a characterization of SCR  irreducible factors  of $x^n - 1$  and  presented a recursive formula for the number of such factors.



In this paper,  we  focus on SR  and SCR  irreducible factors of $x^n-\lambda$, where $\lambda$ is a nonzero element. In Section 2, we deal with   SR irreducible factors. First, we give
a characterization of SR irreducible polynomials. Second,  we obtain an enumeration formula of  SR irreducible factors of $x^n-\lambda$ over $\Bbb F_{q}$. Final, we apply these results to count the number of Euclidean LCD constacyclic codes and prove some known results on Euclidean self-dual constacyclic codes in a simple and direct way.
In Section 3, we deal with SCR irreducible factors of $x^n-\lambda$ over $\Bbb F_{q^2}$. In Section 4, we conclude this paper.

For convenience, we introduce the following notations in this paper.

\begin{tabular}{ll}
$q$ & a power of a prime $p$,   \\

$\Bbb F_{q}$ & finite field $GF(q)$,\\

$\lambda$ & a nonzero element of order $r$ in  $\Bbb F_{q}$,\\



$\phi(n)$& the  Euler's totient function, i.e., the number of positive integers $i$, \\ 
& $1\le i\le n$, with $\gcd(i,n)=1$, \\
$rad(n) $ & the product of distinct prime divisors of a given integer $n$,\\

$\ord_{n} (m)$ &  the order of $m$ modulo $n$ if $\gcd(m,n)=1$,\\

$v_{p}(n)$ & the maximum number $k$ such that $p^k\mid n$ and $p^{k+1}\nmid n$,\\


$f^{\ast}(x)$ & the reciprocal polynomial of the polynomial $f(x)$,\\
$f^{\dagger}(x)$ & the conjugate reciprocal polynomial of the polynomial  $f(x)$,\\

$N_1$ & the number of  irreducible factors of $x^n-\lambda$ over $\Bbb F_q$,\\
$N_2$ & the number of SR irreducible factors of $x^n-\lambda$ over $\Bbb F_q$ ,\\
$M_1$ & the number of  irreducible factors of $x^n-\lambda$ over $\Bbb F_{q^2}$,\\
$M_2$ & the number of SCR irreducible factors of $x^n-\lambda$ over $\Bbb F_{q^2}$.\\

\end{tabular}



\section{Self-reciprocal polynomials}

In this section, we deal with SR polynomials over finite fields. 

\subsection{A general characterization of self-reciprocal polynomials}$~$


\begin{defn}\cite[Definition 3.12]{L}  Let $\Bbb F_q$ be a finite field and $$f(x)=x^n+a_{n-1}x^{n-1}+\cdots+a_1x+a_0\in \Bbb F_q[x]$$ with $a_0\neq0$. Then the reciprocal polynomial $f^{*}(x)$ of $f(x)$ is defined by
$$f^{*}(x)=a_0^{-1}x^nf(\frac{1}{x})=x^n+a_0^{-1}a_1x^{n-1}+\cdots+a_0^{-1}a_{n-1}x+a_0^{-1}.$$ Moreover, $f(x)$ is called self-reciprocal (SR)  if $f(x)=f^{*}(x)$.
\end{defn}
\begin{lem} {\rm Let $s$ be a positive integer, $t\ge 2$ a positive integer,  and $l$  an odd prime. If there exists a positive integer $w$ such that $s^{w}\equiv -1\pmod l$, then there  exists a positive integer $w'$ such that $s^{w'}\equiv -1 \pmod {l^t}$ and $v_2(w)=v_2(w')$. In fact, we can take $w'=wl^{t-1}$. }
\end{lem}
\begin{proof} Let $s^w\equiv -1\pmod {l}$. Then $l\mid s^w+1$. 
Since 
 $\frac{s^{wl}+1}{s^w+1}=\sum_{i=0}^{l-1}(-s^w)^i=l+\sum_{i=0}^{l-1}((-s^w)^i-1)$, $l\mid \frac{s^{wl}+1}{s^w+1}$ and
   $v_l(s^{wl}+1)\ge v_l(s^{w}+1)+1$. By induction,  $v_l(s^{wl^t}+1) \ge v_l(s^w+1)+t$. Hence we can take $w'=wl^{t-1}$ such that $s^{w'}\equiv -1\pmod {l^t}$ and  $v_2(w')= v_2(w)$.
This completes the proof.
\end{proof}

\begin{thm}{\rm Let $n = 2^{v_2(n)}p_{1}^{\alpha_{1}}p_{2}^{\alpha_{2}} \cdots p_{l}^{\alpha_{l}}$ be  the prime factorization, where $p_1,\ldots, p_l$ are distinct odd primes and $\alpha_1, \ldots, \alpha_l$ are positive integers. Let  $\Bbb F_{q}$ be a finite field with $\gcd(n,q)=1$. Suppose that  $f(x)$ is an irreducible factor of $\Phi_n(x)$ over $\Bbb F_{q}$, where $\Phi_n(x)$ is the cyclotomic polynomial of order $n$.  Then the following  are equivalent:

(1) The polynomial $f(x)$ is SR.

(2) There exists a positive integer $w$ such that $q^{w}\equiv -1\pmod n$.

(3) $v_2(d_1)=\cdots=v_2(d_l)=\delta$ and $q\equiv -1\pmod {2^{v_2(n)}}$, where $d_i$ are the order of $q$ modulo $p_{i}$,  $1\le i\le l$, and $\delta$ is a positive integer satisfying $$\delta\left\{\begin{array}{ll}
>0, &\mbox{ if $v_2(n)\le 1$,}\\
=1, &\mbox{ if $v_2(n)\ge 2$.
}\end{array}\right.$$






  }

\end{thm}

\begin{proof}


$(1) \iff (2)$.  Let $\mathbb{Z}_n$ be the ring of integers modulo $n$. For convenience, let $\mathbb{Z}_n =\{0,1,2,\ldots , n - 1\}$.
For an element $s \in \mathbb{Z}_n$, the $q$-cyclotomic coset of $s$ modulo $n$ is
\begin{equation*}C_s=\{s,sq, sq^2,\ldots, sq^{(l_s-1)}\} \mbox{ mod } n \subseteq \mathbb{Z}_n,\end{equation*} where $l_s$ is the smallest positive integer such that $s\equiv sq^{l_s}\pmod n$ and $|C_s|=l_s$.  Let $m=\ord_n(q)$ and  $\alpha$  a generator of $\Bbb F_{q^{m}}^{\ast}$. Then $\beta=\alpha^{\frac{q^m-1}{n}}$ is an element of order $n$ in $\Bbb F_{q^{m}}$.  Suppose that  $f(x) $ is the minimal polynomial of $\beta^s$ over $\Bbb F_q$, where $\gcd(s,n)=1$  and $\deg(f(x))=l$. Then the $q$-cyclotomic coset  corresponding to $f(x)$ is  \begin{equation*}
C_s=\{s,sq, sq^2,\ldots, sq^{(l-1)}\}.\end{equation*}
By the definition,   the $q$-cyclotomic coset  corresponding to the reciprocal polynomial of $f(x)$  is \begin{equation*}
C_{-s}=\{-s,-sq, -sq^2,\ldots, -sq^{(l-1)}\}.\end{equation*} Hence $f(x)$
  is SR if and only if $C_s=C_{-s}$ if and only if there exists a positive integer $w$ such that $q^{w}\equiv -1\pmod n$.

  $(2) \Longrightarrow (3)$. Suppose that  there exists a positive integer $w$ such that $q^{w}\equiv -1\pmod n$. Then
  \begin{eqnarray*}
\begin{cases}
q^{w}\equiv -1\pmod {2^{v_2(n)}},\\
q^{w}\equiv -1\pmod {p_1},\\
\cdots\cdots\\
q^{w}\equiv -1\pmod {p_l}.\\
   \end{cases}
\end{eqnarray*}
Let  $d_i=\ord_{p_i}(q)$, $i=1,\ldots, l$.  Then 
 there exists a positive integer $w$ such that $q^w\equiv{-1}\pmod {p_i}$ if and only if $d_i$ is even and $v_2(d_i)=v_2(w)+1$. Hence  $v_2(w)+1=v_2(d_1)=\cdots=v_2(d_l)=\delta>0$. If $v_2(n)\le 1$, then 
  $q\equiv -1\pmod {2^{v_2(n)}}$  by $\gcd(n,q)=1$. If $v_2(n)\ge 2$, then $q^w\equiv -1\pmod 4$. Hence $w$ is odd and $\delta =1$. Since $q$ is odd and \begin{eqnarray*}q^w+1
  &=&(q+1)(q^{w-1}-q^{w-2}+\cdots+q^2-q+1),\end{eqnarray*}
   $v_2(q^w+1)=v_2(q+1)$ and  $q^{}\equiv -1\pmod {2^{v_2(n)}}$.

  $(3)  \Longrightarrow (2) $. If  $v_2(d_1)=\cdots=v_2(d_l)=\delta>0$, then by Lemma 2.2, there exist odd integers $w_1', \cdots, w_l'$ such that
  \begin{eqnarray*}
\begin{cases}
q^{2^{\delta-1}w_1'}\equiv -1\pmod {p_1^{\alpha_1}},\\
\cdots\cdots\\
q^{2^{\delta-1}w_l'}\equiv -1\pmod {p_l^{\alpha_l}}.\\
   \end{cases}
\end{eqnarray*}
Then  $q^{2^{\delta-1}w_1'w_2'\cdots w_l'}\equiv -1\pmod {p_1^{\alpha_1}\cdots p_l^{\alpha_l}}$. Next we divide the proof into three cases.

(i) If $v_2(n)=0$, i.e., $n$ is odd, then $w= 2^{\delta-1}w_1'w_2'\cdots w_l'$ yields the result.

(ii)
If $v_2(n)=1$, then $q$ is odd as $\gcd(n,q)=1$. Then
   $q^{2^{\delta-1}w_1'w_2'\cdots w_l'}\equiv -1\pmod {2p_1^{\alpha_1}\cdots p_l^{\alpha_l}}$.

  (iii) If $v_2(n)\ge 2$, then  $v_2(d_1)=\cdots=v_2(d_l)=\delta=1$, $q\equiv -1\pmod {2^{v_2(n)}}$, and 
  $w= \frac{1}{2^l}d_1\cdots d_l p_1^{\alpha_1-1}\cdots p_l^{\alpha_l-1}$ is odd. Hence  $q^w\equiv -1\pmod {2^{v_2(n)}}$ and  $q^w\equiv -1\pmod n$.

 This completes the proof.\end{proof}

\subsection{Enumeration formulas}$~$

   In this subsection, we will give some enumeration formulas of numbers of irreducible factors and SR irreducible factors of $x^n-\lambda$ over $\Bbb F_q$, where $\lambda$ is a nonzero element of order $r$ in $\Bbb F_q$.

 Suppose that $\zeta_{nr}$ is a primitive $nr$-th root of unity in an extension field of $\Bbb F_q$,  $\lambda=\zeta_{nr}^n$, and $s$ is a positive integer with  $\gcd(s, r)=1$.
 Then \begin{eqnarray*}x^n-\lambda^s
=x^n-\zeta_{nr}^{sn}
=\prod_{i=0}^{n-1}(x-\zeta_{nr}^{s+ri}).
\end{eqnarray*}
The multiset
\begin{equation}R_s(n,r)=\bigg\{\bigg\{ \ord(\zeta_{nr}^{s+ri}):0\le i\le n-1\bigg\}\bigg\}\end{equation}
is called
 {\it the root distribution} of $x^n-\lambda^s$.

  \begin{thm} {\rm Suppose that $\lambda$ is a nonzero element of order  $r$ in $\Bbb F_q$. Let $n=n_1n_2$ be a positive  integer,  $\gcd(n,q)=1$,  $rad(n_1)\mid rad(r)$, and $\gcd(n_2, r)=1$.  Then for all integers $s$,   $1\le s\le r$, $\gcd(s,r)=1$,
$$R_s(n, r)=\{\{(n_1\phi({d}))\times(rdn_1): d\mid n_2\}\},$$
where $(n_1\phi({d}))\times(rdn_1)$ means that $rdn_1$ repeats $n_1\phi(d)$ times in the multiset.   }
  \end{thm}

   \begin{proof} By $x^n-\lambda^s
=\prod_{i=0}^{n-1}(x-\zeta_{nr}^{s+ri})$,
 the order of each $\zeta_{nr}^{s+ri}$ is $$\frac{nr}{\gcd(s+ri,nr)}=r\cdot \frac{n}{\gcd(s+ri,n)}.$$
Then the root distribution of $x^n-\lambda^s$ is given by the following multiset

$$R_s(n,r)=\bigg\{\bigg\{ r\times\frac{n}{\gcd(s+ri,n)}:0\le i\le n-1\bigg\}\bigg\}.$$

To prove the result, it is suffice to determine the multiset $$\{\{\gcd(s+ri,n):0\le i\le n-1\}\}$$
 for any integer $s$ with $1\le s\le r$ and $\gcd(s,r)=1$.

Let $n=n_1n_2$, where $rad(n_1) \mid rad(r)$ and $\gcd(n_2,r)=1$. Since $\gcd(s,r)=1$, $\gcd(s+ri,n_1)=1$. Then by $\gcd(r,n_2)=1$ and $r\Bbb Z_{n_2}=\Bbb Z_{n_2}$,
\begin{eqnarray*}&&\{\{\gcd(s+ri,n):0\le i\le n-1\}\}\\
&=&\{\{\gcd(s+ri,n_1n_2):0\le i\le n-1\}\}\\
&=&\{\{\gcd(s+ri,n_2):cn_2\le i\le (c+1)n_2-1, 0\le c\le n_1-1\}\}\\
&=&\{\{n_1\times\gcd(s+ri,n_2):0\le i\le n_2-1\}\}\\
&=&\{\{n_1\times\gcd(s+i,n_2):0\le i\le n_2-1\}\}\\
&=&\{\{n_1\times\gcd(i,n_2):0\le i\le n_2-1\}\}\\
&=&\{\{(n_1\phi(\frac{n_2}{d}))\times d:d\mid n_2\}\}.
\end{eqnarray*}
The result follows from $\frac{rn}{d}=rn_1\cdot \frac{n_2}d$.
 \end{proof}

\begin{thm} {\rm Suppose that $\lambda$ is a nonzero element of order  $r$ in $\Bbb F_q$. Let $n=n_1n_2$ be a positive  integer, $\gcd(n,q)=1$,   $rad(n_1) \mid rad(r)$,  and $\gcd(n_2,r)=1$.


$(1)$ Then  \begin{equation*}  n=\frac{1}{\phi(r)}\sum_{d| n_2}\phi{(rdn_1)}=\sum_{d|n_2}n_1\phi(d). \end{equation*}

$(2)$ The number  of  irreducible factors of $x^n-\lambda$ over $\Bbb F_q$ is
\begin{equation*}N_{1}=\frac{1}{\phi(r)}\sum_{d| n_2}\frac{\phi{(rdn_1)}}{\ord_{rdn_1}(q)}=n_1\sum_{d| n_2}\frac{\phi{(d)}}{\ord_{rdn_1}(q)}.\end{equation*}

$(3)$ The number  of self-reciprocal irreducible factors of $x^n-\lambda$ over $\Bbb F_q$ is
\begin{equation}N_{2}=n_1\sum_{d\in S(n,r)}\frac{\phi{(d)}}{\ord_{rdn_1}(q)},\end{equation}
where \begin{eqnarray*}S(n,r)=\bigg\{ d\mid n_2:
\begin{array}{l}
\mbox{there exists a positive integer $w$}\\
\mbox{such that $q^w\equiv-1\pmod {rdn_1}$}
\end{array}\
\bigg\}.\end{eqnarray*}
}

\end{thm}
\begin{proof} (1) By the process of  proving Theorem 2.4, the  root  distribution  of  $x^n-\lambda^s$ is $\{\{n_1\phi(d)\times rdn_1: d\mid n_2\}\}$.
By $rad(n_1)\mid rad(r)$, $$\phi(n_1r)=n_1r\prod_{prime\ p|n_1r}(1-\frac 1{p})=n_1r\prod_{prime\ p|r}(1-\frac 1p)=n_1\phi(r). $$
By
$\gcd(n_2, rn_1)=1$,  \begin{eqnarray*} &&\sum_{d| n_2}\phi{(rn_1d)}=\phi{(rn_1)}\sum_{d|n_2}\phi(d)=\phi(r)n_1n_2=\phi(r)n.
\end{eqnarray*}

(2)  Note that $$\Phi_r(x^n)=\prod_{\mbox{\tiny$
\begin{array}{c}
s=1\\
\gcd(s,r)=1\\
\end{array}$
}}^r(x^n-\lambda^s).$$
The degree of $\Phi_r(x^n)$ is $n\phi(r)$.  By Theorem 2.4 and    $\Phi_{rdn_1}(x)\mid \Phi_r(x^n)$ for each $d\mid n_2$,   the number of irreducible factors of $x^n-\lambda$ over $\Bbb F_q$ is \begin{eqnarray*}N_1&=&\frac{1}{\phi(r)}\sum_{d| n_2}\frac{\phi{(rn_1d)}}{\ord_{rdn_1}(q)}\\
&=&\frac{1}{\phi(r)}\sum_{d| n_2}\frac{\phi(rn_1)\phi(d)}{\ord_{rdn_1}(q)}\\
&=&n_1\sum_{d| n_2}\frac{\phi(d)}{\ord_{rdn_1}(q)}.
\end{eqnarray*}


(3) The result follows from (2) and Theorem 2.3.

This completes the proof.
\end{proof}

We give two examples to illustrate Theorem 2.5.

\begin{exa} {\rm Let $q=7$ and $n=27$. Then the number of irreducible factor and the number of self-reciprocal irreducible factors of $x^n-\lambda$ over $\Bbb F_q$ are given in Table 1. These results are confirmed by Magma.

\[ \begin{tabular} {c} Table $1$.  Example $2.6$\\
\begin{tabular}{|c|c|c|c|c|c|c|c|}
  \hline
Order of $\lambda$&$\{rdn_1:d\mid n_2\}$&$N_1$& $N_2$\\
\hline
$1$&$\{1,3,9,27\}$&$7$&$ 1$\\
 \hline
 $2$&$\{1,3,9,27\}$ &$7$&$ 1$
\\\hline
$3$&$\{81\}$&$1$&$0$  \\
\hline
$6$&$\{162\}$&$1$ & $ 0$ \\
\hline
\end{tabular}
\end{tabular}
\]

}

\end{exa}

\begin{exa}  {\rm Let $q=19$ and $n=36$. Then the number of irreducible factor and the number of self-reciprocal irreducible factors of $x^n-\lambda$ over $\Bbb F_q$ are given in Table 2. These results are confirmed by Magma.

\[ \begin{tabular} {c} Table $2$.  Example $2.7$\\
\begin{tabular}{|c|c|c|c|c|c|c|c|}
  \hline
Order of $\lambda$&$\{rdn_1:d\mid n_2\}$& $N_1$&  $N_2$\\
\hline
$1$&$\{1,2,3,4,6,9,12,18,36\}$&$27$& $4$\\
 \hline
 $2$&$\{8, 24,72\}$ &$18$& $0$
\\\hline
$3$&$\{27, 54, 108\}$&$9$&$0$  \\
\hline
$6$&$\{216\}$&$6$ & $ 0$ \\
\hline
$9$&$\{81, 162,324\}$& $3$& $0$  \\
\hline
$18$&$\{648\}$&$2$&$0$  \\
\hline
\end{tabular}
\end{tabular}
\]
}
\end{exa}

\subsection{Euclidean LCD and self-dual constacyclic codes}$~$




For two vectors $x=(x_0,x_1,\ldots, x_{n-1})$ and $y=(y_0,y_1,\ldots, y_{n-1}) \in \Bbb F_q^n$, the Euclidean inner product is defined as $\langle x,y\rangle_E=\sum_{i=0}^{n-1}x_iy_i$. For a $\lambda$-constacyclic code $\mathcal{C}$ of length $n$ over $\Bbb F_q$, the {\it Euclidean dual} code of $\mathcal{C}$ is defined as $$\mathcal{C}^{\bot_E}=\{u\in \Bbb F_q^n|\langle u,v\rangle_E=0 \mbox{ for all } v\in \mathcal{C}\}.$$ The code $\mathcal{C}$ is called {\it Euclidean LCD}   if $\mathcal{C}\cap\mathcal{C}^{\bot_E}=\{0\}$ and {\it Euclidean self-dual} if $\mathcal{C}=\mathcal{C}^{\bot_E}$.  Assume that $\mathcal{C}=\left\langle g(x)\right\rangle$ is a $\lambda$-constacyclic code length $n$ over $\Bbb F_q$.  As we know, the generator polynomial of   $\mathcal{C}^{\bot_E}$ is $h^*(x)$,  where  $h^*(x)$ is   the reciprocal polynomial of the polynomial $h(x)=\frac{x^n-\lambda}{g(x)}$.

\begin{lem}  {\rm Let $\Bbb F_q$ be a finite field and $\lambda\in \Bbb F_q^*$.  Let $n$ be a positive integer and $\gcd(n,q)=1$. Suppose that $\mathcal{C}$ is a $\lambda$-constacyclic code and
$x^n-\lambda$ has the following irreducible factorization over $\Bbb F_q$:
$$x^n-\lambda=e_1(x)\cdots e_u(x)f_1(x)f_1^*(x)\cdots f_v(x)f_v^*(x),$$ where $u,v$ are nonnegative integers, $e_i(x), 1\le i\le u,$ are SR, and  $f_j^*(x)$ are the reciprocal polynomial of $f_j(x)$, $1\le j\le v$. Then

$(1)$ $\mathcal{C}=\left\langle g(x)\right\rangle$ is an Euclidean LCD code if and only if $g(x) $ is SR i.e., $g(x)$ has the following form:  $$g(x)=e_1^{\alpha_1}(x)\cdots e^{\alpha_u}_u(x){f_1(x)}^{\beta_1}{f_1^*(x)}^{\beta_1}\cdots {f_v(x)}^{\beta_v}{f_v^*(x)}^{\beta_v}, $$ where $\alpha_i,\beta_j\in \{0,1\}, 1\le i\le u$, $1\le j\le v$.

$(2)$ $\mathcal{C}=\left\langle g(x)\right\rangle$ is Euclidean self-dual if and only if  $u=0$ and $g(x)$ has the following form: $${f_1(x)}^{\beta_1}{f_1^*(x)}^{1-\beta_1}\cdots {f_v(x)}^{\beta_v}{f_v^*(x)}^{1-\beta_v},$$ where  $ \beta_j\in \{0,1\}$, $1\le j\le v$.}
\end{lem}

\begin{proof}  If $\mathcal{C}=\left\langle g(x)\right\rangle$ is a $\lambda$-constacyclic code of length $n$ over $\Bbb F_q$, then $\mathcal{C}^{\bot_E}=\langle h^*(x)\rangle $,  where  $h(x)=\frac{x^n-\lambda}{g(x)}$. Then $\mathcal{C}$ is Euclidean LCD if and only if $\mbox{lcm}(g(x),  h^*(x))=x^n-\lambda$ and
$\mathcal{C}$ is Euclidean self-dual if and only if $g(x)=h^*(x)$.
This completes the proof.
\end{proof}




For the case of Euclidean LCD constacyclic codes, it was shown that any $\lambda$-constacyclic code with $\lambda \notin\{-1,1\}$ is an LCD code  in \cite{D5}. For any $\lambda\in \Bbb F_q^*$,   $x^n-\lambda$ is SR if and only if $\lambda^2=1$.


\begin{thm}[Euclidean LCD constacyclic codes]   {\rm Let $\Bbb F_q$ be a finite field.  Let $n$ be a positive integer and $\gcd(n,q)=1$.  Then there are three cases.

$(1) $ If $\lambda\notin\{-1,1\}$, then any $\lambda$-constacyclic code is an Euclidean LCD code.

$(2)$
There are $2^{\frac{n+N_2}2}$ Euclidean LCD cyclic codes,  where $N_2$ is given in Eq. (2.2).

$(3)$ There are $2^{\frac{n+N_2}2}$ Euclidean LCD negacyclic codes, where  $N_2$ is given in Eq. (2.2).
}
\end{thm}

\begin{proof} The results follow from Theorem 2.5 and Lemma 2.8 (1).
\end{proof}


\begin{thm}[Euclidean  self-dual constacyclic codes]  {\rm Let $\Bbb F_q$ be a finite field.  Let $n= 2^{v_2(n)}n'$ be a positive integer, $v_2(n)>0$,   and $\gcd(n,q)=1$.

$(1)$  There is no Euclidean self-dual cyclic codes of length $n$ over $\Bbb F_q$.

$(2)$ Euclidean self-dual negacyclic codes of length $n$ over $\Bbb F_q$  exist if and only if $q\not\equiv-1 \pmod {2^{v_2(n)+1}}$. }
\end{thm}


\begin{proof} (1) Since $x-1$ is always  SR irreducible factor  of $x^n-1$ over $\Bbb F_q$,  by Lemma 2.8 (2) the result holds.


(2) By Theorem 2.5,  $q$ is odd, $r=2$, $n_1=2^{v_2(n)}$,  $n_2=n'$, and   
\begin{eqnarray*}S(n,2)=\bigg\{ d\mid n':
\begin{array}{l}
\mbox{there exists a positive integer $w$}\\
\mbox{such that $q^w\equiv-1\pmod {2^{v_2(n)+1}d}$.}
\end{array}\
\bigg\}.\end{eqnarray*}

If   Euclidean self-dual negacyclic codes of length $n$ over $\Bbb F_q$ exist, then every irreducible factor of $\Phi_{2^{v_2(n)+1}d}(x)$ is not SR over $\Bbb F_q$ for each $d\mid n'$, i.e.  $ S(n,2)=\emptyset$.  Hence $q\not\equiv-1 \pmod {2^{v_2(n)+1}}$.

Conversely, 
if there is no Euclidean self-dual negacyclic codes of length $n$ over $\Bbb F_q$,  then there exists some $d\in S(n,2)$  and some positive integer $w$ such that $q^w\equiv -1\pmod  {2^{v_2(n)+1}d}$ by Theorem 2.3, i.e. $S(n,2)\ne \emptyset$. Hence $q^w\equiv -1\pmod {2^{v_2(n)+1}}$.  By the proof of Theorem 2.3, $q\equiv -1\pmod {2^{v_2(n)+1}}$.

This completes the proof.
\end{proof}

\begin{rem} 
{\rm
Theorem 2.10~ (2) has been proved in \cite[Theorem 3]{B}. 

}
\end{rem}




\section{Self-conjugate-reciprocal polynomials}

In this section, we deal with self-conjugate-reciprocal polynomials over finite fields. 
\subsection{A general characterization of self-conjugate-reciprocal polynomials}$~$

\begin{defn}\cite{BJU}  Let $\Bbb F_{q^2}$ be a finite field and $$f(x)=x^n+a_{n-1}x^{n-1}+\cdots+a_1x+a_0\in \Bbb F_{q^2}[x]$$ with $a_0\neq0$. Then the conjugate-reciprocal polynomial $f^{\dagger}(x)$ of $f(x)$ is defined by
$$f^{\dagger}(x)=x^n+a_0^{-q}a_1^qx^{n-1}+\cdots+a_0^{-q}a_{n-1}^qx+a_0^{-q}.$$
 Moreover, $f(x)$ is called self-conjugate-reciprocal (SCR)   if $f(x)=f^{\dagger}(x)$.
\end{defn}





\begin{thm}{\rm Let $n =2^{v_2(n)}p_{1}^{\alpha_{1}}p_{2}^{\alpha_{2}} \cdots p_{l}^{\alpha_{l}}$ be  the prime factorization, where $p_1,\ldots, p_l$ are distinct odd primes and $\alpha_1, \ldots, \alpha_l$ are positive integers. Let  $\Bbb F_{q^2}$ be a finite field and $\gcd(n,q)=1$. Suppose that  $f(x)$ is an irreducible factor of $\Phi_n(x)$ over $\Bbb F_{q^2}$.  Then  the following are equivalent:

$(1)$ The   polynomial $f(x)$ is SCR.

$(2)$  There exists an odd positive integer $w$ such that $q^{w}\equiv -1\pmod n$.

$(3)$ $q\equiv -1\pmod {2^{v_2{(n)}}}$ and $v_2(d_1)=\cdots=v_2(d_l)=1$, where $d_i$ is the order of $q$ modulo $p_{i}$, $1\le i\le l$.}





\end{thm}

\begin{proof}
(1) $\iff$ (2). Suppose that  $f(x) $ is the minimal polynomial of $\beta^s$ over $\Bbb F_{q^2}$, where  $\gcd(s,n)=1$  and $\deg(f(x))=l$. Then the $q$-cyclotomic coset  corresponding to $f(x)$ is \begin{equation*}C_s=\{s,sq^2, sq^4,\ldots, s(q^2)^{(l-1)}\}.\end{equation*}
Similar to the proof of Theorem 2.3,    the $q^2$-cyclotomic coset  corresponding to the conjugate-reciprocal polynomial of $f(x)$  is \begin{equation*}C_{-sq}=\{-sq,-sq^3,\ldots, -sq(q^2)^{(l-1)}\}.\end{equation*} Hence $f(x)$
  is SCR if and only if $C_s=C_{-sq}$ if and only if  there exists a positive integer $k$ such that $s(q^2)^{k}\equiv -sq\pmod n$ if and only if there exists an odd positive integer $w$ such that $q^{w}\equiv -1\pmod n$.


(2) $\iff$ (3).  This result follows from the proof of Theorem 2.3.

  This completes the proof.
\end{proof}

\subsection{Enumeration formulas}$~$


By Theorem 3.2, we have the following result analogous to  Theorem 2.5.
\begin{thm} {\rm Suppose that $\lambda$ is a nonzero element of order  $r$ in $\Bbb F_{q^2}$. Let $n=n_1n_2$ be a positive  integer, $\gcd(n,q)=1$, $rad(n_1) \mid rad(r)$, and $\gcd(n_2,r)=1$. Then



$(1)$ The number  of  irreducible factors of $x^n-\lambda$ over $\Bbb F_{q^2}$ is
\begin{equation*}M_{1}=n_1\sum_{d| n_2}\frac{\phi{(d)}}{\ord_{rdn_1}({q^2})}.\end{equation*}

$(2)$ The number  of self-reciprocal irreducible factors of $x^n-\lambda$ over $\Bbb F_{q^2}$ is
\begin{equation}M_{2}=n_1\sum_{d\in T(n,r)}\frac{\phi{(d)}}{\ord_{rdn_1}({q^2})},\end{equation}
where \begin{eqnarray*}T(n,r)=\bigg\{ d\mid n_2:
\begin{array}{l}
\mbox{there exists an odd positive integer $w$}\\
\mbox{such that $q^w\equiv-1\pmod {rdn_1}$}
\end{array}\
\bigg\}.\end{eqnarray*}
}





\end{thm}

If $\lambda=1$, then \cite[Theorem 2.22] {BJU} is an immediate consequence of Theorem 3.3.

\begin{cor} {\rm Let $q$ be a prime power and let $l_1, l_2, \ldots, l_t$ be distinct odd primes relatively prime to $q$. For each $1\le i\le t$, let $r_i$ be a positive integer and let $\ord_{l_i} (q)=2^{a_i}b_i$, where $a_i\ge 0$
 is an integer and $b_i \ge 1$ is an odd integer.  

$(1)$ If there exists $j\in \{1,2,\ldots, t\}$ such that $a_j=0$ or $a_j\ge 2$, then $$T(\prod_{i=1}^tl_i^{r_i},1)=T(\prod_{i=1}^{j-1}l_i^{r_i}  \prod_{i=j+1}^{t}l_i^{r_i} ,1).$$

$(2)$ If $a_1=a_2=\cdots=a_t=1$ and $n=\prod_{i=1}^tl_i^{r_i}$, then the number  of SCR irreducible factors of $x^n-1$ over $\Bbb F_{q^2}$ is
\begin{equation*}M_{2}=\sum_{d\mid n}\frac{\phi{(d)}}{\ord_{d}(q^2)}.\end{equation*}

$(3)$ If $a_i\neq 1$ for all $i\in \{1,2,\ldots, t\}$ and $n=\prod_{i=1}^tl_i^{r_i}$, then the only SCR irreducible factors of $x^n-1$ over
$\Bbb F_{q^2}$ is $x-1$.}
\end{cor}

We give two examples to illustrate Theorem 3.3.

\begin{exa}{\rm Let $q=16$ and $n=27$. Then the number of irreducible factors and the number of  self-reciprocal irreducible factors of $x^n-\lambda$ over $\Bbb F_q$ are given in Table 3. These results are confirmed by Magma.

\[ \begin{tabular} {c} Table $3$.  Example $3.5$\\
\begin{tabular}{|c|c|c|c|c|c|c|c|}
 \hline
Order of $\lambda$&$\{rdn_1:d\mid n_2\}$&$M_1$& $M_2$\\
\hline
$1$&$\{1,3,9,27\}$&$7$& $1$\\
 \hline
 $3$&$\{81\}$ &$1$&$ 0$
\\\hline
$5$&$\{5, 15, 45, 135\}$&$7$&$0$  \\
\hline
$15$&$\{405\}$&$1$ & $ 0$ \\
\hline\end{tabular}
\end{tabular}
\]
}
\end{exa}

\begin{exa}{\rm Let $q=25$ and $n=36$. Then the number of irreducible factors and the number of  self-reciprocal irreducible factors of $x^n-\lambda$ over $\Bbb F_q$ are given in Table 4. These results are confirmed by Magma.

\[ \begin{tabular} {c} Table $4$.  Example $3.6$\\
\begin{tabular}{|c|c|c|c|c|c|c|c|}
 \hline
Order of $\lambda$&$\{rdn_1:d\mid n_2\}$&$M_1$& $M_2$\\
\hline
$1$&$\{1,2,3,4,6,9,12,18,36\}$&$20$& $2$\\
 \hline
 $2$&$\{8, 24,72\}$ &$20$&$ 0$
\\\hline
$3$&$\{27, 54, 108\}$&$4$&$0$  \\
\hline
$4$&$\{16, 48,144\}$&$10$ & $ 0$ \\
\hline
$6$&$\{216\}$& $4$& $0$  \\
\hline
$8$&$\{32, 96, 288\}$&$6$&$0$  \\
\hline
$12$&$\{432\}$&$2$&$0$  \\
\hline
$24$&$\{864\}$&$1$&$0$  \\
\hline\end{tabular}
\end{tabular}
\]

}

\end{exa}

\subsection{Hermitian LCD and self-dual constacyclic codes}$~$

For two vectors $x=(x_0,x_1,\ldots, x_{n-1})$ and $y=(y_0,y_1,\ldots, y_{n-1}) \in \Bbb F_{q^2}^n$, the Hermitian inner product is defined as $\langle x,y\rangle_H=\sum_{i=0}^{n-1}x_iy_i^q$. For a $\lambda$-constacyclic code $\mathcal{C}$ of length $n$ over $\Bbb F_{q^2}$, the {\it Hermitian dual} code of $\mathcal{C}$ is defined as $$\mathcal{C}^{\bot_H}=\{u\in \Bbb F_{q^2}^n|\langle u,v\rangle_H=0 \mbox{ for all } v\in \mathcal{C}\}.$$ The code $\mathcal{C}$ is called {\it Hermitian LCD}   if $\mathcal{C}\cap\mathcal{C}^{\bot_H}=\{0\}$ and {\it Hermitian self-dual} if $\mathcal{C}=\mathcal{C}^{\bot_H}$.  If  $\mathcal{C}=\left\langle g(x)\right\rangle$ is a $\lambda$-constacyclic code of length $n$ over $\Bbb F_{q^2}$, then the generator polynomial of $\mathcal{C}^{\bot_H}$ is $h^{\dagger}(x)$, where $h^{\dagger}(x)$ is the conjugate-reciprocal polynomial of  $h(x)=\frac{x^n-\lambda}{g(x)}$.





We have the following result analogous to Lemma 2.8, 

\begin{lem}{\rm  Let $\Bbb F_{q^2}$ be a finite field and $\lambda\in \Bbb F_{q^2}^*$.  Let $n$ be a positive integer and $\gcd(n,q)=1$. Suppose that $\mathcal{C}$ is a $\lambda$-constacyclic code and
$x^n-\lambda$ has the following irreducible factorization over $\Bbb F_{q^2}$:
$$x^n-\lambda=e_1(x)\cdots e_u(x)f_1(x)(f_1^{\dagger}(x))\cdots f_v(x)(f_v^{\dagger}(x)),$$ where $u,v$ are nonnegative integers, $e_i(x), 1\le i\le u,$ are SCR, and  $f_j^*(x)$ are the conjugate-reciprocal polynomial of $f_j(x)$, $1\le j\le v$. Then

$(1)$ $\mathcal{C}=\left\langle g(x)\right\rangle$ is a Hermitian LCD code if and only if $g(x) $ is SCR i.e., $g(x)$ has the following form:  $$g(x)=e_1^{\alpha_1}(x)\cdots e^{\alpha_u}_u(x){f_1(x)}^{\beta_1}{f_1^{\dagger}(x)}^{\beta_1}\cdots {f_v(x)}^{\beta_v}{f_v^{\dagger}(x)}^{\beta_v}, $$ where $\alpha_i,\beta_j\in \{0,1\}, 1\le i\le u$, $1\le j\le v$.

$(2)$ $\mathcal{C}=\left\langle g(x)\right\rangle$ is Hermitian self-dual if and only if  $u=0$ and $g(x)$ has the following form: $${f_1(x)}^{\beta_1}{f_1^{\dagger}(x)}^{1-\beta_1}\cdots {f_v(x)}^{\beta_v}{f_v^{\dagger}(x)}^{1-\beta_v},$$ where  $ \beta_j\in \{0,1\}$, $1\le j\le v$.

}



\end{lem}


For the case of Hermitian LCD constacyclic codes, it was shown that any $\lambda$-constacyclic code over $\Bbb F_{q^2}$ with $\lambda^{1+q} \neq 1$ is an LCD code in  \cite[Corollary 3.3]{LFL}. For any $\lambda\in \Bbb F_{q^2}^*$,   $x^n-\lambda$ is SCR if and only if $\lambda^{q+1}=1$.


By Theorem 3.3 and Lemma 3.7 (1), we have the following theorem.

\begin{thm}[Hermitian LCD constacyclic codes] {\rm
 Let $\Bbb F_{q^2}$ be a finite field.  Let $n$ be a positive integer and $\gcd(n,q)=1$.

$(1)$ If $\lambda^{q+1}\neq 1$, then any $\lambda$-constacyclic code is a Hermitian LCD code.

$(2)$   If $\lambda^{q+1}=1$, then there are  $2^{\frac{n+M_2}2}$ Hermitian LCD $\lambda$-constacyclic codes, where $M_2$ is given in Eq. (3.1).}
\end{thm}

\begin{thm}[Hermitian self-dual constacyclic codes]
{\rm 
 Let $ \Bbb F_{q^2}$ be a finite field of odd order $q^2$. 
 Let  $\lambda $ a nonzero element of order $r=2^{v_2(r)}r'$ in $ \Bbb F_{q^2}$ and   $r\mid(q+1)$.   Let $n=2^{v_2(n)}n'=n_1n_2$ be a positive integer, $v_2(n)>0$,  $\gcd(n,q)=1$, $rad(n_1) \mid rad(r)$, and $\gcd(n_2,r)=1$.
Then Hermitian self-dual $\lambda$-constacyclic codes of length $n$ over $\Bbb F_{q^2}$  exist if and only if $v_2(r)>0$ and $q \not\equiv-1 \pmod {2^{v_2(n)+v_2(r)}}$.}
\end{thm}

\begin{proof} 

By Lemma 3.7 (2) and Theorem 3.2, Hermitian self-dual $\lambda$-constacyclic codes of length $n$ over $\Bbb F_{q^2}$  exist if and only if for any $d\mid n_2$, $q^w\not\equiv-1\pmod {rdn_1}$ for every odd integer $w$.

If $v_2(r)=0$, i.e., $r$ is odd, then $n_1$ is odd and $rad(rn_1)\mid (q+1)$. By Lemma 2.2, there exists a positive integer $w$ such that $q^w\equiv -1\pmod {rn_1}$, which is a  contradiction. Therefore we have $v_2(r)>0$.

If $r$ is even, then $v_2(rdn_1)=v_2(n_1)+v_2(r)= v_2(n)+v_2(r)$ for each $d\mid n_2$. Then $q^w\not\equiv-1\pmod {rdn_1}$ implies that $q\not\equiv-1\pmod  {2^{v_2(n)+v_2(r)}}$.

On the other hand, if $r$ is even and $q\not\equiv-1\pmod  {2^{v_2(n)+v_2(r)}}$, then there are no irreducible SCR polynomial of order $rdn_1, d\mid n_2$ over $\Bbb F_{q^2}$ as $2^{v_2(n)+v_2(r)} \mid rdn_1$. By Lemma 3.7, there exist Hermitian self-dual $\lambda$-constacyclic codes of length $n$ over $\Bbb F_{q^2}$.



This completes the proof.
\end{proof}

\begin{rem} {\rm Theorem $3.9$ has been proved in \cite[Theorem 3.9]{YC}. 
}
\end{rem}

\section{Concluding remarks}
In this paper, we presented some general characterizations on  self-reciprocal (SR) and  self-conjugate-reciprocal (SCR) irreducible polynomials over finite fields. By using these characterizations, we obtained some enumeration formulas of  SR  and  SCR irreducible factors of $x^n-\lambda$ over finite fields; counted the numbers of Euclidean  and Hermitian LCD constacyclic codes; proved some well-known results on Euclidean  and Hermitian self-dual constacyclic codes by a simple  and direct way.

Recently, many scholars investigated Galois constacyclic codes over finite fields, see \cite{CMTQ, FZ,LL,LM, LP, LYH, LFL}. Motivated by their work, we introduce the definition of {\bf $\sigma$-self-reciprocal polynomials} over finite fields. Let $\Bbb F_q$ be a finite field of order $q$ and  $\sigma $ be an automorphism  over  some  subfield of $\Bbb F_q$. Then the $\sigma$-reciprocal polynomial  of $f(x)\in \Bbb F_q[x]$ is defined by $\sigma(f^{*}(x))$. Moreover, $f(x)$ is called $\sigma$-self-reciprocal polynomial if $f(x)=\sigma(f^{*}(x))$. It may be interesting to investigate some properties and applications of these $\sigma$-self-reciprocal polynomials. 

\section*{Acknowledgments}
Part of this work was done when the first author was visiting Korea Institute for Advanced Study (KIAS),  Seoul, South Korea. Y. Wu would like to thank the institution for the kind hospitality. This paper was supported by the National Natural Science Foundation of China under Grant 61772015 and  the   Foundation of Science and Technology on Information Assurance Laboratory under Grant KJ-17-010.  




\begin{thebibliography}{[kkk]}

\bibitem{BJU} A. Boripan, S. Jitman, P. Udomkavanich, Self-conjugate-reciprocal irreducible monic factors of $x^n-1$ over finite fields and their applications, Finite Fields Appl., 55: 78-96, 2019.



\bibitem{B}  T. Blackford,  Negacyclic duadic codes,  Finite Fields Appl.,  14(4): 930-943, 2008.

\bibitem{CMTQ} C. Carlet, S. Mesnager, C. Tang, Y. Qi, On $\sigma$-LCD codes, IEEE Trans. Inf. Theory, 65(3): 1694-1704, 2019.


\bibitem{CDL}  B. Chen, H. Q. Dinh, H. Liu,  Repeated-root constacyclic codes of length $lp^s$ and their duals, Discrete
Appl. Math., 177: 60-70,  2014.


  \bibitem{CDL2} B. Chen, H. Q. Dinh, H. Liu, Repeated-root constacyclic codes of length $2l^mp^n$, Finite Fields Appl.,
33: 137-159, 2015.

\bibitem{CDLW}  B. Chen, H. Q. Dinh, H. Liu, L. Wang,  Constacyclic codes of length $2p^s$ over $\Bbb F_{p^m}+ u\Bbb F_{p^m}$,  Finite Fields Appl., 37: 108-130, 2016.


  \bibitem{CFLL}  B. Chen, Y. Fan, L. Lin, H. Liu, Constacyclic codes over finite fields, Finite Fields Appl., 18(6):
 1217-1231, 2012.

\bibitem{CLZ}   B. Chen, S. Ling, G. Zhang,   Enumeration formulas for self-dual cyclic codes, Finite Fields Appl.,  42:  1-22, 2016.

  \bibitem{D1} H. Q. Dinh, Repeated-root constacyclic codes of length $2p^s$, Finite Fields Appl., 18:
133-143, 2012.


  \bibitem{D3} H. Q. Dinh, Structure of repeated-root constacyclic codes of length $3p^s$ and their duals, Discrete
Math., 313(9): 983-991, 2013.



  \bibitem{D5} H. Q. Dinh, Repeated-root cyclic and negacyclic codes of length $6p^s$, AMS Contemp. Math., 609: 69-87, 2014.



     \bibitem{D7}  H.Q. Dinh, H. D. T. Nguyen, S. Sriboonchitta, and T. M. Vo, Repeated- root constacyclic codes of prime power lengths over finite chain rings, Finite Fields Appl.,  43: 22-41,  2017.

\bibitem{F} N. Fernando,  Self-reciprocal polynomials and coterm polynomials,  Des. Codes Cryptogr.,  86(8):  1707-1726, 2018.



\bibitem{FR} N. Fernando, M. Rashid, Fibonacci self-reciprocal polynomials and Fibonacci permutation polynomials, arXiv:1712.07723, 2018.

\bibitem{FZ} Y. Fan, L. Zhang, Galois self-dual constacyclic codes, Des. Codes Cryptogr., 84: 473-492, 2017.

\bibitem{G1} T. Garefalakis,   Self-reciprocal irreducible polynomials with prescribed coefficients,  Finite Fields Appl.,  17(2): 183-193, 2011.


\bibitem{GK} T. Garefalakis , G. Kapetanakis  On the Hansen-Mullen conjecture for self-reciprocal irreducible polynomials, Finite Fields Appl.,  18(4): 832-841, 2012.








\bibitem{G2} T. A. Gulliver,  Self-reciprocal polynomials and generalized Fermat numbers,  IEEE Trans. Inf. Theory,  38(3):  1149-1154, 1992.


\bibitem{GB} T. A. Gulliver, V. K. Bhargava,  Some properties of self-reciprocal polynomials,  Appl. Math. Lett.,  3(3): 47-51, 1990.

\bibitem{HB}  S. Hong, D. Bossen,  On some properties of self-reciprocal polynomials (Corresp.),  IEEE Trans. Inf. Theory,  21(4):  462-464, 1975.



 \bibitem{JLX} Y. Jia, S. Ling, C. Xing, On self-dual cyclic codes over finite fields, IEEE Trans. Inf. Theory, 57:  2243-2251, 2011.



\bibitem{L1} C. Li, Hermitian LCD codes from cyclic codes,  Des. Codes Cryptogr.,  86: 2261-2278, 2018.




  \bibitem{LDL} C. Li, C. Ding, S. Li, LCD cyclic codes over finite fields, IEEE Trans. Inform. Theory, 63(7): 4344-4356, 2017.


\bibitem{L} R. Lidl, H. Niederreiter,   Finite fields, Cambridge University Press, 1997.

  \bibitem{LL} H. Liu, J. Liu, $\sigma $-self-orthogonal constacyclic codes of length $ p^ s $ over $\mathbb F_ {p^ m}+ u\mathbb F_ {p^ m} $, arXiv preprint arXiv:1807.09474, 2018.


  \bibitem{LM} H. Liu, Y. Maouche,  Some repeated-root constacyclic codes over Galois rings,  IEEE Trans. Inf. Theory, 63 (10): 6247-6255, 2017.

 \bibitem{LP} H. Liu, X. Pan,  Galois hulls of linear codes over finite fields, Des. Codes Cryptogr., 2019. https://doi.org/10.1007/s10623-019-00681-2.
 
 
 \bibitem{LLKZ} L. Liu, L. Li, X. Kai, et al. Repeated-root constacyclic codes of length $3lp^s$ and their dual codes, Finite Fields Appl., 42: 269-295, 2016.



\bibitem{LYH}  X. Liu, L. Yu, P. Hu,  New entanglement-assisted quantum codes from $k$-Galois dual codes, Finite Fields Appl., 55: 21-32, 2019.

 \bibitem{LFL} X. Liu, Y. Fan, H. Liu,  Galois LCD codes over finite fields,   Finite Fields Appl., 49: 227-242, 2018.







\bibitem{M} J. W. Massey, Reversible codes, Inf. Contr., 7: 369-380, 1964.











\bibitem{WY}  Y. Wu, Q. Yue,  Factorizations of binomial polynomials and enumerations of LCD and self-dual constacyclic codes,  IEEE Trans. Inf. Theory, 65(3): 1740-1751, 2019.

\bibitem{YM1}  X. Yang, J. L. Massey, The condition for a cyclic code to have a complementary dual, Discrete Math., 126: 391-393, 1994.

\bibitem{YC} Y. Yang, W. Cai,  On self-dual constacyclic codes over finite fields, Des. Codes Cryptogr.,  74(2): 355-364, 2015.

\bibitem{YM2} J. L. Yucas, G. L. Mullen,  Self-reciprocal irreducible polynomials over finite fields, Des. Codes Cryptogr.,  33(3): 275-281, 2004.


\bibitem{ZTG} W. Zhao, X. Tang, Z. Gu, Constacyclic  codes  of  length  $kl^{m}p^{n}$ over  a finite field, Finite Fields Appl., 52:  51-66, 2018.


   \end{thebibliography}
\end{document}